\documentclass[aps,prb,twocolumn,floatfix,amsmath,amssymb,showpacs,
               superscriptaddress]{revtex4-1}
\usepackage{graphicx,placeins}
\usepackage{float}
\usepackage{color}
\usepackage{dcolumn}
\usepackage{bm}
\usepackage{epsfig,psfrag,amsmath,amssymb,float}
\input{epsf}
\usepackage{hyperref}
\usepackage[percent]{overpic}
\hypersetup{
    colorlinks=true,
    linkcolor=blue,
    citecolor=blue,
    filecolor=magenta,      
    urlcolor=cyan,
}
\usepackage{array}
\newcolumntype{C}[1]{>{\centering\arraybackslash$}p{#1}<{$}}
\usepackage{titlesec}

\setcitestyle{square}

\bibpunct{[}{]}{,}{n}{}{}
\begin{document}
\title{Magnetic ground states of honeycomb lattice Wigner crystals}

\author{Nitin Kaushal}
\affiliation{Materials Science and Technology Division, Oak Ridge National 
Laboratory, Oak Ridge, Tennessee 37831, USA}

\author{Nicol\'as Morales-Dur\'an}
\affiliation{Department of Physics, The University of Texas at Austin, Austin, Texas 78712, USA}

\author{Allan H. MacDonald}
\affiliation{Department of Physics, The University of Texas at Austin, Austin, Texas 78712, USA}


\author{Elbio Dagotto}
\affiliation{Materials Science and Technology Division, Oak Ridge National 
Laboratory, Oak Ridge, Tennessee 37831, USA}
\affiliation{Department of Physics and Astronomy, The University of 
Tennessee, Knoxville, Tennessee 37996, USA}

\date{\today}

\begin{abstract}
Lattice Wigner crystal states stablilized by long-range Coulomb interactions
have recently been realized in two-dimensional 
moir\'e materials. We employ large-scale unrestricted Hartree-Fock techniques
to unveil the magnetic phase diagrams of honeycomb lattice Wigner crystals.
For the three lattice filling factors with the largest charge gaps, $n=2/3, 1/2, 1/3$,
the magnetic phase diagrams contain multiple phases, including ones with non-collinear and non-coplanar spin 
arrangements. We discuss magnetization evolution with external magnetic field, which has 
potential as an experimental signature of exotic spin states.  Our theoretical results could potentially be
validated in moir\'e materials formed from group VI transition metal dichalcogenide twisted homobilayers.
\end{abstract}
\maketitle

\section{Introduction}
Overlaying two-dimensional crystal layers with small lattice constant mismatches 
or interlayer twist angles creates moir\'e patterns.
Recent experimental progress~\cite{CZhang01,YPan01,Tang01,Regan01,McGilly01,TLi01} has 
established transition metal dichalcogenide (TMD) bilayers with moir\'e patterns periods on the 
10 nm scale as an attractive platform~\cite{Wu01,Wu02} to synthesize artificial one-band 
Hubbard model strongly-correlated electron systems.  The minibands of these moir\'e materials are 
flat for a wide-range of twist angles, unlike twisted bilayer graphene which requires tuning to magic-angles~\cite{Bistritzer01, Cao01, Cao02,Andrei02}. 
Key control parameters of these artificial Hubbard models
like the bandwidth, the carrier density, and the screening of 
long-range Coulomb interactions can be adjusted by varying twist-angle, gate voltage, 
and gate electrode placement~\cite{Kennes01,Andrei01}.

Because they are based on triangular lattice host atomic crystals~\cite{Xiao01}, TMD moir\'e materials 
realize either triangular lattice Hubbard models or, when the emergent low-energy 
moir\'e Hamiltonian has $C_6$ symmetry, honeycomb lattices.
For example, the topmost moir\'e band of aligned $\textrm{WSe}_{2}/\textrm{WS}_{2}$ heterobilayers and twisted $\textrm{WSe}_{2}$ homobilayers~\cite{Zhang01,Wang01, Ghiotto01}
mimics the triangular lattice Hubbard model, while the topmost moir\'e bands of $\textrm{MoSe}_{2}$, $\textrm{MoS}_{2}$, and $\textrm{WS}_{2}$ homobilayers have been predicted to simulate the honeycomb lattice Hubbard model~\cite{Angeli01,Xian01,Vitale01}.  It has been found that in moir\'e materials,
interactions lead not only to Mott insulating states 
at half-filling~\cite{Chu01} (hole band filling $n=1$), but also at fractional fillings~\cite{Huang01,Xu01,Jin01}.
The ground states at fractional filling factors are Wigner crystal states in which 
electrons occupy a subset of the available lattice sites and translational symmetry is broken, as shown for example by recent high-resolution scanning tunneling experiments for hole fillings $n=2/3$, $1/2$, and $1/3$ in $\textrm{WSe}_{2}/\textrm{WS}_{2}$ ~\cite{HLi01}. Wigner crystal states rely on inter-site Coulomb interactions,
and are likely obscured by disorder in heavily doped atomic crystals. Their prominence in experiment 
is a demonstration of the potential of moir\'e materials to 
reveal new physics. Although Wigner crystal states have charge gaps, their low-energy physics is non-trivial because of the 
spin-degrees of freedom that remain on all occupied lattice sites.  The spins are expected to order at low-temperatures in most cases, but could potentially have spin-liquid ground states in some instances.
The interactions between spins in Wigner crystals states are different from those in Mott insulator 
states because the possibilities for virtual electron hopping across the charge gap are enriched, and control
the magnetic ground states that are the 
main subject of this paper. %

\begin{figure}[!h]
\hspace*{-0.5cm}
\vspace*{0cm}
\begin{overpic}[width=1.0\columnwidth]{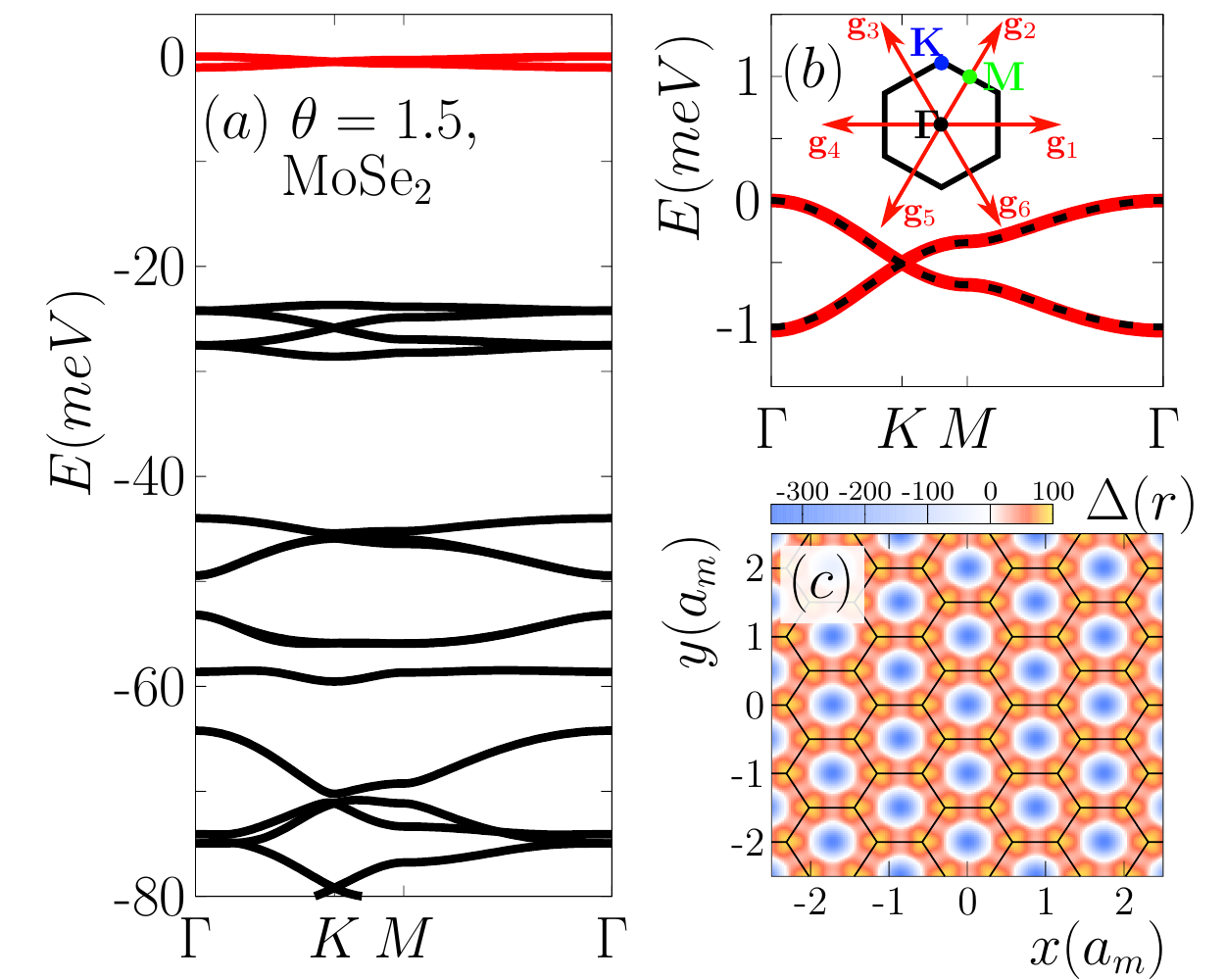}
\end{overpic}
\caption{Panel (a) shows a nearly flat $\textrm{MoSe}_{2}$ band structure calculated 
using the continuum model in Ref.~\cite{Angeli01} at twist angle $\theta=1.5^{\circ}$.
In panel (b), the topmost bands are magnified. The dashed black line is the band structure of a 
nearest-neighbour tight binding model defined on a honeycomb lattice. Panel (c) shows the moir\'e potential in real space. The model parameters $(V_{1},V_{2},V_{3},\phi_{s},m^{*},a_{0})$=$(36.8, 8.4, 10.2, \pi, 1.17m_{e}, 3.295{\rm \AA})$ are fixed to values corresponding to $\textrm{MoSe}_{2}$.  Here $m_{e}$ is the bare electron mass 
and $V_{1,2,3}$ are in ${\rm meV}$ units.}
\label{fig0_intro}
\end{figure}
To date the triangular lattice Hubbard model has been the main focus for both 
experimental and theoretical studies of TMD moir\'e materials~\cite{Duran01,Hu01}. However, $\textrm{MoSe}_{2}$, $\textrm{MoS}_{2}$, and $\textrm{WS}_{2}$ homobilayers with the $2\textrm{H}$ structure have
$\Gamma$ point~\cite{Manzeli01} interlayer antibonding states at the valence band maxima (VBM), unlike $\textrm{MoTe}_{2}$ and $\textrm{WSe}_2$ where the VBM is located at the $K$ point. This has important
implications for the moir\'e bands of these materials.  Recent {\it ab-initio} calculations showed that a small twist angle ($\theta$) in $\Gamma$-valley TMD homobilayers leads to the creation of 
moir\'e bands mimicking one and two orbitals  honeycomb lattices, as well as kagome lattices~\cite{Angeli01,Xian01,Vitale01}. The continuum model Hamiltonian was derived in~\cite{Angeli01} by keeping only antibonding layer states and neglecting spin-orbit coupling because it vanishes at the $\Gamma$ point.  

For concreteness, we specifically discuss the example of the $\textrm{MoSe}_{2}$ homobilayer with twist angle $\theta=1.5^{\circ}$. The continuum moir\'e Hamiltonian is ${H} = -\hbar^{2} k^{2}/2m^{*} + 
\Delta(\mathbf{r})$. In this equation, $\Delta(\mathbf{r})$ is the moir\'e potential 
defined as 
$
\Delta(\mathbf{r}) = \sum_{s}\sum_{j=1}^{6}V_{s}e^{i(\mathbf{g}_{j}^{s}\cdot\mathbf{r} + \phi_{s})},
$
where $\mathbf{g}_{j}^{s}$ are vectors of the moir\'e reciprocal lattice connecting 
to the $s$-th nearest-neighbour site. 
The moir\'e Brillouin zone and the $\mathbf{g}_{j}^{1}$ vectors are shown in the inset of Fig.~\ref{fig0_intro}(b). In Fig.~\ref{fig0_intro}(a), the band structure for $\textrm{MoSe}_{2}$ is displayed. 
The topmost red colored bands, highlighted in Fig.~\ref{fig0_intro}(b), realize 
a one-orbital honeycomb lattice tight-binding model.  The dashed black line shows
the nearest-neighbour tight-binding model result with hopping parameter $t=1/6$~meV. 
This comparison demonstrates that for small values of the twist angle $\theta$, the highest energy bands can be faithfully described by the nearest-neighbour honeycomb lattice. The emergence of a honeycomb lattice is understood intuitively by noticing the structure of the moir\'e potential, plotted in Fig.~\ref{fig0_intro}(c), in which the moir\'e potential maxima define a honeycomb lattice structure. 

Once established that the $\Gamma$-valley bilayer-TMDs can effectively be described 
as a honeycomb lattice, it is anticipated that in their Mott insulator state (hole filling $n=1$), simple collinear bipartite antiferromagnetic order is supported. However at the fractional fillings, the open questions arise: (i) Do we, as in the triangular lattice heterobilayer 
case, obtain 2D generalized Wigner crystals at special fillings in these moir\'e materials once the model is made
realistic by including longer range electronic interactions?
(ii) What kind of magnetic states we may expect in these Wigner crystals? 
The present work investigates the above questions by obtaining the many-body ground state for different particle fillings of the honeycomb lattice, considering the strong on-site Coulomb interaction $U_{0}>>t$ limit, and incorporating long-range Coulomb interactions as well. The longer range interactions are essential to understand the phases of the fractionally filled bilayer-TMDs. However, investigating fractionally-filled Hubbard models with non-local interactions is a formidable task, even for exact Lanczos studies on small clusters or highly accurate density matrix renormalization group studies on ribbons. For this reason static mean-field theory investigations have been extensively employed by the community to study moir\'e superlattices~\cite{Zang01,Pan01,Pan02,Pan03,Pan04}. In this publication, we will employ an unrestricted Hartree-Fock approximation to exhaustively 
explore the phase diagrams of long-range Coulomb interacting Hubbard models on honeycomb lattices, with focus on fillings $n=2/3$, $1/2$, and $1/3$.  In insulating states, the unrestricted Hartree-Fock approximation 
allows energy to be minimized by varying the spin-direction on each lattice site and 
is therefore equivalent to employing a classical approximation within a spin-only model for the 
low-energy physics.  The static mean field theory approximation is accurate when  
charge fluctuations are suppressed, as they are in Wigner Crystal states, and ground state 
spin-fluctuations are also weak.  Importantly the Hartree-Fock completely eliminates self-interaction
effects in the limit of electrons localized on lattice sites. However, the Hartree-Fock approximation cannot 
describe spin-liquid states.  

Using the Hartree-Fock approximation we are able to address how the ground state evolves as the parameters of the model are varied. After establishing the phase diagrams and the presence of rich spin physics in the fractionally filled honeycomb moir\'e materials, in the last portion of our work we also discuss the magnetization evolution under external magnetic fields for the magnetic states in the strong long-range Coulomb interaction limit. These magnetization {\it vs.} external magnetic field curves are experimentally accessible, and 
can be used as signatures of the exotic magnetic Wigner crystals unveiled here.
 
\section{Lattice Hamiltonian}\label{Hamil}

In this study, we extend the one-band Hubbard model on a honeycomb lattice by
including the long-range Coulomb repulsion that is relevant in moir\'{e} systems. The Hamiltonian is
\begin{multline}\label{H_RealSpace}
H=\sum_{i, i^{\prime}, a,b, \sigma}t_{i^{\prime} b}^{ia}c_{ia\sigma}^{\dagger}c_{i^{\prime}b\sigma}^{\phantom{\dagger}} + U_{0}\sum_{i,a}n_{ia\uparrow}n_{ia\downarrow} \\+ U_{1}\sum_{i}n_{i0}n_{i1} + 1/2\sum_{i\ne i^{\prime}} U_{ab}^{ii^{\prime}}n_{ia}n_{i^{\prime}b},
\end{multline}
where $c_{i a \sigma}^{\dagger} (c_{i a \sigma}^{\phantom{\dagger}})$ is the standard fermionic creation (anhilation) operator. In the Hamiltonian above, $\{i,i^{\prime}\}$ labels unit cell, ${a,b}=0/1$ labels sublattice, 
and $\sigma \in \{\uparrow, \downarrow\}$ denotes spin. $n_{i a \sigma} = c_{i a \sigma}^{\dagger}c_{i a \sigma}^{\phantom{\dagger}}$ is the fermionic density operator. The first term of Eq.~\ref{H_RealSpace} captures the band Hamiltonian whose hopping parameters we choose as the unit of energy by setting 
$t_{i^{\prime}b}^{i a}=t=-1$ for nearest neighbors and to $0$ otherwise. 
We kept only the nearest-neighbour hopping because for $\Gamma$-valley moir\'{e} 
homobilayers the longer range hoppings are negligible compared with $|t|$, as shown in the supplementary~\cite{supple} material. Moreover, we showed specifically for $\textrm{MoSe}_{2}$ homobilayers that topmost bands can be well represented by just a nearest-neighbour honeycomb lattice (See Fig.~\ref{fig0_intro}(b).) The second ($U_{0}$) and third ($U_{1}$) terms represent the on-site and the nearest-neighbour Coulomb repulsion, respectively. The fourth term is the long-range tail of the Coulomb interaction, where the parameters $U_{ab}^{ii^{\prime}}$ are fixed using the screened Coulomb repulsion $U(\mathbf{r})=U_{1}(\frac{1}{|\mathbf{r}|} - \frac{1}{\sqrt{|\mathbf{r}|^{2} + d^{2}}})/(\frac{1}{|\mathbf{r}_{1}|} - \frac{1}{\sqrt{|\mathbf{r}_{1}|^{2} + d^{2}}})$, where $|\mathbf{r}_{1}|$ is the nearest-neighbor distance, $|\mathbf{r}|$ is the distance between sites $\{i,a\}$ and $\{i^{\prime},b\}$, and $d$ is the screening length. We choose $U_1$ and $U_0$ as independent model parameters
because $U_0$ is sensitive mainly to twist angle and the strength of the moir\'e modulation potential, 
which control the size of the band Wannier function \cite{Wu01}, whereas $U_1$ is sensitive mainly to the screening background of the moir\'e material.  We expect the ratio of the strength of the long-range 
Coulomb tail to $U_1$ to be nearly universal, as we have assumed. We solved the above model using the unrestricted Hartree-Fock approximation with details provided
in the Methods section.

\begin{figure}[!h]
\hspace*{-0.5cm}
\vspace*{0cm}
\begin{overpic}[width=0.75\columnwidth]{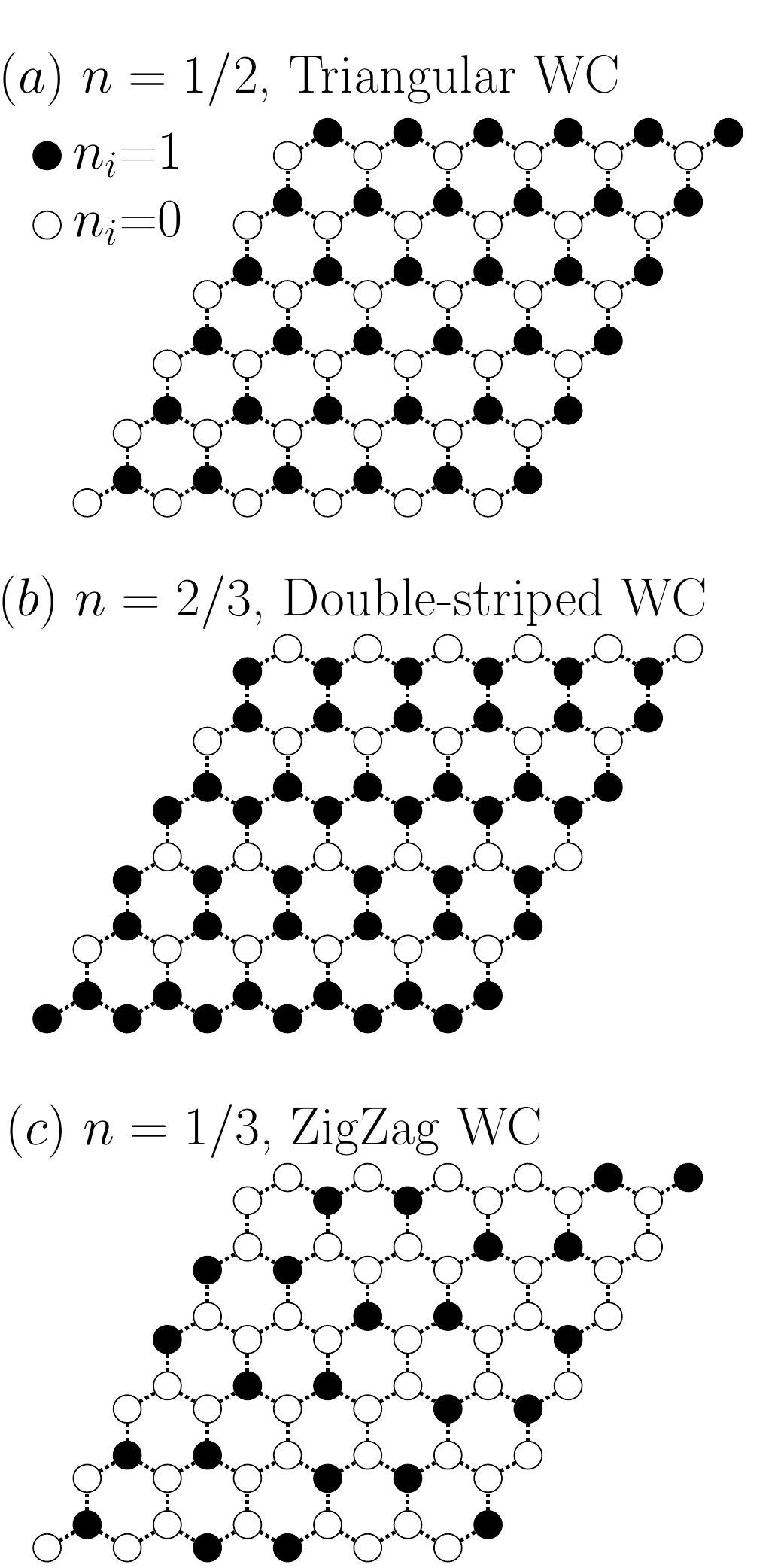}
\end{overpic}
\caption{Wigner crystals in the limit of $U_{0},U_{1}>>t$ for the fillings $n=1/2$, $n=2/3$, and $n=1/3$ are shown in panels (a), (b), and (c), respectively.}
\label{fig0p5}
\end{figure}

\begin{figure*}[!ht]
\hspace*{-0.5cm}
\vspace*{0cm}
\begin{overpic}[width=2.2\columnwidth]{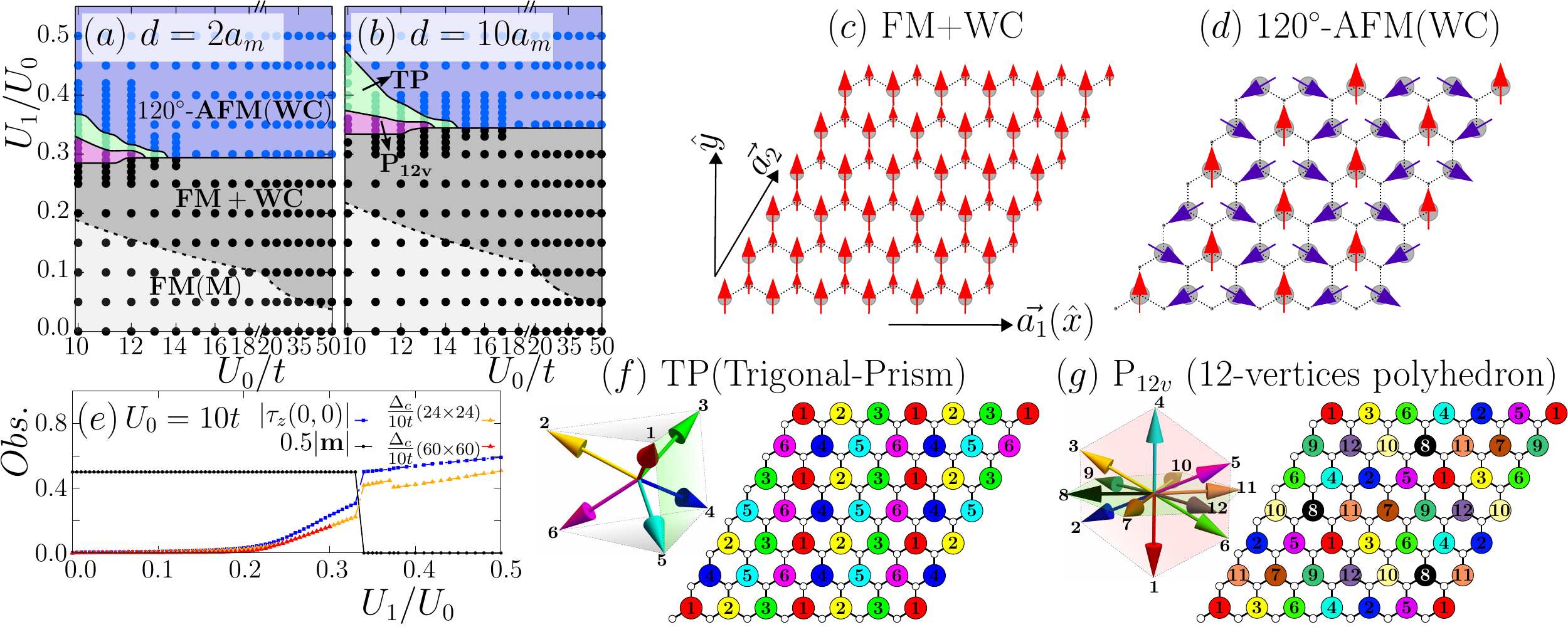}
\end{overpic}
\caption{Results for filling $n=1/2$. In (a) and (b), the $U_{1}/U_{0}$ vs $U_{0}/t$ phase diagrams are shown for the screening lengths $d=2a_{m}$ and $d=10a_{m}$, respectively. Panels (c) and (d) illustrate the FM+WC (Ferromagnetic
Wigner Crystal) and $120^{\circ}\textrm{-AFM}$ states respectively. 
Panels (f) and (g) show the non-coplanar states and the respective polyhedra corresponding to phases $\textrm{TP}$ (Trigonal-Prism) and $\textrm{P}_{12v}$ (12-vertices polyhedron), respectively. The size of the arrows and circles in panels (c,d,f,g) is proportional to the magnitude of the local spin $\langle \mathbf{S}_{i} \rangle$ size and local density $\langle n_{i}\rangle$, respectively. Panel (e) shows the net magnetizaion $|\mathbf{m}|$ and the pseudospin structure factor $\tau_{z}(\mathbf{q})$ at momentum $\mathbf{q}=(0,0)$ for ${24\times 24}$ system size, and the charge gap $\Delta_{c}$ for the ${24\times24}$ and $60\times60$ system sizes, for various values of $U_{1}/U_{0}$, at fixed $U_{0}/t=10$ and $d=10a_{m}$.}
\label{fig1}
\end{figure*}

\section{Results at $n$=1/2, 2/3, 1/3}

In the Hamiltonian used here, once the screening length $d$ is fixed 
the only free parameters are $U_{0}$, $U_{1}$, and the band filling $n$.
We explored the complete $U_{1}/U_{0}$ vs $U_{0}/t$ phase diagrams, fixing system size to $12(L_{x})\times12(L_{y})$, for the average electronic densities $n=1/2,2/3,1/3$, where $n=N_{e}/(2L_{x}L_{y})$, $N_{e}$ is the total number of electrons and $L_{x}(L_{y})$ is number of unit cells along the $x(y)$ direction. The model we study has particle-hole
symmetry with respect to $n=1$ since we employ only nearest-neighbour hopping, hence the magnetic states discussed in our paper for any fractional filling $n$ will also be present for the case of the filling $2-n$.

When quantum fluctuations are suppressed and we are deep in the Wigner crystal (WC) regime we obtain the configurations shown in Fig.~\ref{fig0p5} for the fillings addressed. The WC for $n=1/2$ resembles a half-filled standard triangular lattice~\cite{Chern01}. The other two states for $n=1/3$ and $n=2/3$ break rotational symmetry of the honeycomb lattice. In the following, we will calculate the
$U_{0}/t$ {\it vs.} $U_{1}/U_{0}$ phase diagrams of Eq.~\ref{H_RealSpace}. We will mainly discuss the phases appearing in the strong coupling regime, here defined as $U_{0}/t \ge 10$, because 
this is the regime most often explored experimentally in moir\'e superlattices typically and because this is the region where our method performs better. Interestingly we discover a rich set of competing phases for each of the fillings, that vary as the $U_{0}/t$ and $U_1/U_0$ ratios are varied. The details of the phases in the weak and intermediate coupling regions are provided in the supplementary for completeness~\cite{supple}. We also investigated the stability of the various phases found here {\it vs.} changes in screening length, by solving the above model for two screening length values $d=2 a_{m}$ and $10 a_{m}$. The main message of our findings is that honeycomb lattice Wigner crystals 
have complex magnetic states, and that transitions between them can be tuned by varying $U_{0}/t$, which is very sensitive to twist angle and three-dimensional dielectric environment, or by varying $U_{1}/U_{0}$, which is sensitive to both twist angle and metallic screening backgrounds.

\subsection{\it{(a) n=1/2 results}}\label{sec_n1by2}
The main results for filling $n=1/2$ are shown in Fig.~\ref{fig1}. The calculations were performed for all the points shown as small circles in the phase diagrams in Fig.~\ref{fig1}(a) and (b) for screening lengths $d=2\,a_{m}$ and $d=10\,a_{m}$, respectively. Remarkably, we identified multiple phases in the strong coupling limit ($U_{0}>>t$) which are stable for both screening lengths, see Fig.~\ref{fig1}(a,b).

Firstly we will discuss the region of the phase diagram where we found robust triangular WC states (shown in Fig.~\ref{fig0p5}(a)) and Hartee-Fock is most trustworthy as the charge fluctuations are suppressed. 
The triangular WC state appears at large $U_{0}/t$ and large $U_{1}/U_{0}$, since the charge gap requires 
strong on-site and near-neighbor interactions.  We find that the $120^{\circ}$-AFM state, (see Fig.~\ref{fig1}(d)), 
is the most stable magnetic state. In this state the charge density order is identified by the pseudospin structure factor $\tau_{z}(\mathbf{q})=1/N\sum_{\mathbf{i}}\tau_{z}(\mathbf{i})e^{i \mathbf{q}\cdot \mathbf{r}_{\mathbf{i}}}$ at momentum ${\bf{q}}=(0,0)$, where the quantity $\tau_{i,z}=\langle n_{i,b} \rangle-\langle n_{i,a}\rangle$ measures the electronic density difference between the two honeycomb sublattices, indicating the breaking of inversion symmetry between the two sublattices. $\tau_{z}(0,0)$ smoothly decreases upon decreasing $U_{1}/U_{0}$. The total magnetization $|m|=2/N_{e}|\sum_{i}\langle S_{i}\rangle|=0$ throughout this phase.

The $120^{\circ}$-AFM state is common on triangular lattices.  Surprisingly we also found unexpected exotic non-coplanar AFM states: the 12-vertices polyhedron ($\textrm{P}_{12v}$) and the Trigonal-Prism (TP).
Both states have non-zero $\tau_{z}(0,0)$, confirming triangular lattice Wigner crystal charge order. 
Previous to this work, only tetrahedral non-coplanar magnetic states were reported in honeycomb~\cite{Li01} and triangular lattice Hubbard models~\cite{Pasrija01,Martin01}. The $\textrm{P}_{12v}$ phase has 12 distinct spins 
per unit cell with the members of two sets of six spins, namely $\{1,2,3,4,5,6\}$ and $\{7,8,9,10,11,12\}$, lying along two perpendicular hexagons (colored as red and green in Fig.~\ref{fig1}(g)). Figure~\ref{fig1}(g) also shows the positions of the non-coplanar spins in real-space. The Trigonal Prism (TP) phase is shown in Fig.~\ref{fig1}(f) where six spins are aligned towards the corner of the trigonal-prism and the real-space lattice is shown on the right side. Increasing $U_{1}/U_{0}$, inside the region named TP in the phase diagram, decreases the height ($h$ i.e. distance between spins ``1 and 5'' of the prism) and on further increasing $U_{1}/U_{0}$, $h$ suddenly drops to zero, converting the spin-configuration to that of the $120^{\circ}$-AFM state.

\begin{figure*}[!ht]
\hspace*{-0.5cm}
\vspace*{0cm}
\begin{overpic}[width=2.2\columnwidth]{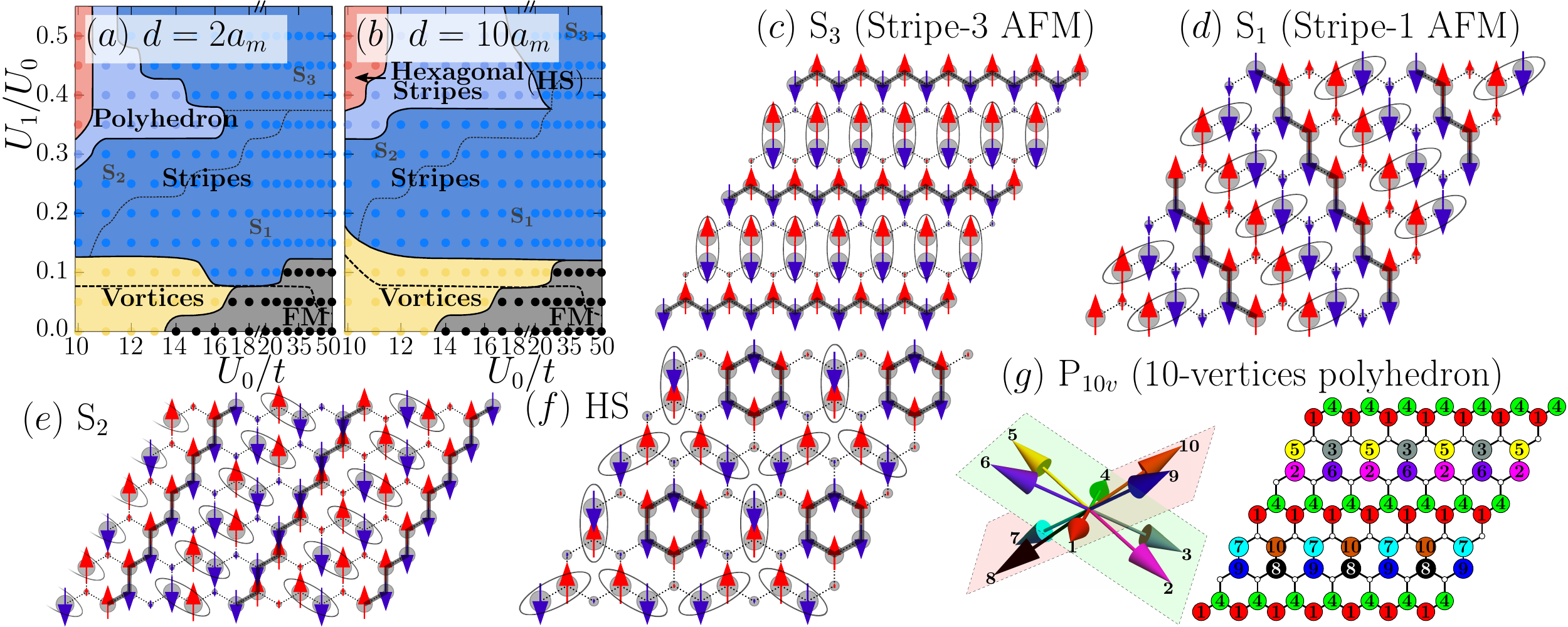}
\end{overpic}
\caption{Results for filling $n=2/3$. In (a) and (b), the $U_{1}/U_{0}$ vs $U_{0}/t$ phase diagrams are shown for screening lengths $d=2a_{m}$ and $d=10a_{m}$, respectively. The metal-insulator transition is illustrated by the thick-dashed black line in the low $U_{1}/U_{0}$ region. Panels (c,d,e,f,g) display representative states of the Stripe-3 AFM ($S_{3}$), Stripe-1 AFM ($S_{1}$), Stripe-2 AFM ($S_{2}$), Hexagonal Stripes $HS$ and 10-vertices polyhedron ($P_{10v}$), respectively. Panel (g) also shows the 10-vertices polyhedron corresponding to the state $P_{10v}$. The set of spins \{1,2,3,4,5,6\} and \{7,8,1,9,10,4\} lie in the green and red planes, respectively.}
\label{fig2}
\end{figure*}

Furthermore, we checked the scalar chirality $\mathbf{S}_{i}\cdot(\mathbf{S}_{j}\times\mathbf{S}_{k})$ of both non-coplanar states, where $\mathbf{S}_{i,j,k}$ are the spins located on the emergent triangular lattice. We find that the chirality has striped patterns, with zero net chirality unlike the tetrahedron spin state with homogenous and nonzero net chirality, for details see~\cite{supple}. Our calculations also confirmed zero Hall-conductance for both non-coplanar states, showing that they are topologically trivial. 

Further decreasing $U_{1}/U_{0}$, we found a sudden decrease in $\tau_{z}(0,0)$ and an abrupt 
jump in $|m|$ from 0 to 1, indicating a first-order AFM to FM phase transition near $U_{1}/U_{0}\sim 0.35$ (for $d=10a_{m}$). This phase also breaks the sub-lattice inversion symmetry, hence making it a saturated FM insulator. A representative state for this FM+WC phase is shown in Fig.~\ref{fig1}(c). In the low $U_{1}/U_{0}$ region, we found a fully polarized ferromagnetic (FM) metal with a net magnetization $|m|=1.0$ i.e. a Nagaoka ferromagnetic state. This fully polarized metallic state has been reported in earlier mean-field studies of the Hubbard models on various two-dimensional lattices, including the honeycomb lattice~\cite{FazekasBook,Hanisch01}.  In Fig.~\ref{fig1}(e) we show $\tau_{z}(0,0)$ and the charge gap $\Delta_{c}$ evolution with changing $U_{1}/U_{0}$, fixing $U_{0}/t=10$ and $d=10a_{m}$. The $\tau_{z}(0,0)$ and $\Delta_{c}$ continuously decrease to $0$, suggesting a second order phase transition in between FM insulator and FM metallic phase. Nevertheless, exact calculations are required to confirm the presence of this FM metallic state as the Hartree-Fock approximation is expected to overestimate the presence of ferromagnetic metallic states,
especially for low band fillings~\cite{Bach01}.

In the triangular Wigner crystal limit, as the charge fluctuations are heavily suppressed, it is natural to ponder if a low-energy spin model can explain the magnetic states we found. At large $U_{1}/U_{0}$, fourth-order processes, using the hopping as a small parameter, generate an antiferromagnetic exchange $J_{AFM}\propto t^4/(2U_{1})^2 U_{0}$ between the nearly half-filled sites of the emergent triangular lattice (which are second-nearest neighbours in the original honeycomb lattice) leading to a spin-1/2 antiferromagnetic Heisenberg model on the triangular lattice. In agreement with this reasoning, we found the $120^{\circ}$-AFM state in the large $U_{1}/U_{0}$ region of the phase diagrams, as in the ground state solution of the triangular lattice Heisenberg model in both classical and quantum limits~\cite{Seabra01,Capriotti01}. 

Interestingly, theoretical studies of classical spin models on triangular lattices, when in presence of a 4-site ring exchange term, have shown the appearance of 6-sublattice and 12-sublattice non-coplanar phases resembling our Trigonal Prism and $\textrm{P}_{12v}$ phases, respectively~\cite{Kubo01}. In our study in the Wigner crystal region, the ring-exchange term on 4-site plaquettes as in triangular lattices can be generated by 8th-order hopping processes leading to a ring exchange $J_{R} \propto t^{8}/(2U_{1})^2(U_{0})^3(3U_{1})^2$. Note that in this scenario  $J_{R}/J_{AFM} \propto t^{4}/(U_{0}\,U_{1})^2$ and decreasing $U_{1}$ increases $J_{R}/J_{AFM}$, showing the importance of the ring-exchange mechanism as $U_{1}$ is decreased. Naturally this suggests that the ring-exchange term will be dominant inside the Wigner crystal phase as both $U_{1}$ and $U_0$ are decreased (if the Wigner crystal remains stable), explaining the location of the non-coplanar phases at the boundary of the Wigner crystal phase and also explaining why the non-coplanar phases are destabilized as $U_{1}$ or $U_{0}$ is increased. 

Quantum fluctuations can be studied using DMRG~\cite{White01,White02}.  Although these 
are limited to finite width ribbons, they will be important to investigate the stability of the non-coplanar states and the possibility of realizing exotic quantum spin liquids. For example, it has been shown that the Trigonal Prism state has uniform vector chirality (although the net scalar chirality is zero).  Including quantum fluctuations may lead to a vector chiral spin liquid state~\cite{Yasuda01}.
\begin{figure*}[!ht]
\hspace*{-0.5cm}
\vspace*{0cm}
\begin{overpic}[width=2.2\columnwidth]{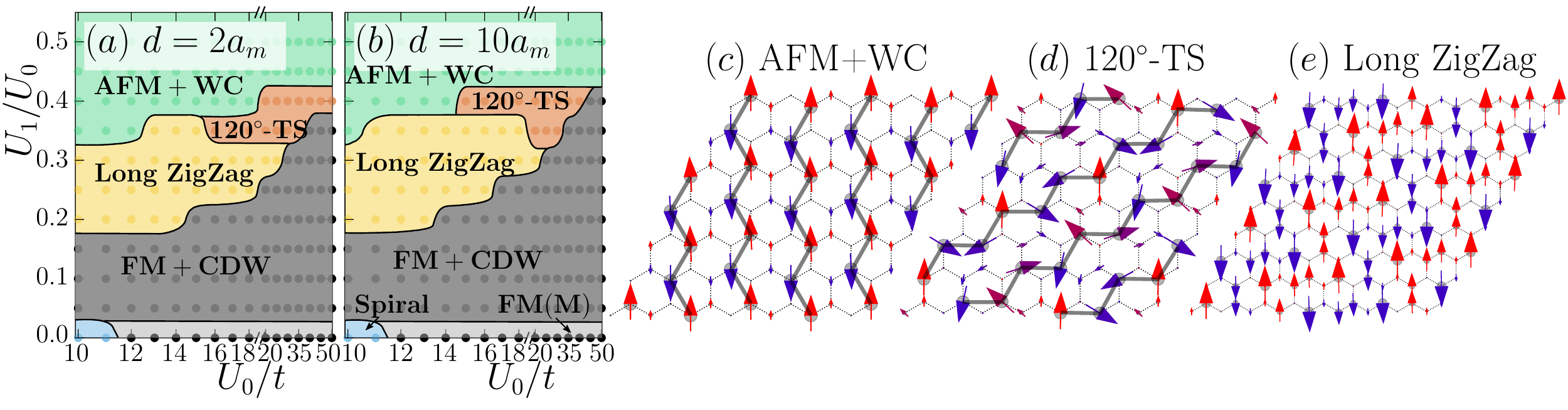}
\end{overpic}
\caption{Results for filling $n=1/3$. In (a) and (b), the $U_{1}/U_{0}$ vs $U_{0}/t$ phase diagrams for screening lengths $d=2a_{m}$ and $d=10a_{m}$ are shown, respectively. Panels (c,d,e) show the representative states of Zigzag Wigner crystal with collinear AFM ordering (AFM+WC), Zigzag $\textrm{120}^{\circ}$-Twin Stripes ($\textrm{120}^{\circ}$-TS), and Long Zigzag collinear phase, respectively.}
\label{fig3}
\end{figure*}

\subsection{\it{(b) n=2/3 results}} 

The $U_{1}/U_{0}$ vs $U_{0}/t$ phase diagrams for filling $n=2/3$ at $d=2a_{m}$ and $d=10a_{m}$ are shown in Fig.~\ref{fig2}(a,b). Again, we unveiled a plethora of phases.  
In a large part of the phase diagram (the blue colored region) we found states with two distinct charge stripes with a width of 2-sites (resembling the double-striped Wigner crystal shown in Fig.~\ref{fig0p5}), where one stripe follows an AFM zigzag pattern and other stripe is made up of nearest neighbour AFM dimers present at distance of $a_{m}$. These double-striped phases break the $C_{3}$ rotational invariance of the honeycomb lattice. Stripe formation in the local 
charge density naturally favours AFM correlations, since antiferromagnetic superexchange is expected 
between the nearest-neighbour nearly half-filled sites. 
We find three kinds of similar looking but distinct collinear antiferromagnetic double-striped states named here $S_{1}$ (see Fig.~\ref{fig2}(d)), $S_{2}$ (see Fig.~\ref{fig2}(e)), and $S_{3}$ (see Fig.~\ref{fig2}(c)). In the $S_{3}$ phase, the orientation of the AFM dimers (shown by ellipses drawn around pair of opposite spins) are parallel to each other in the stripe of dimers; whereas in the $S_{2}$ phase we found AFM dimers are oriented opposite to each other. In the $S_{1}$ phase, which is stabilized for the small $U_{1}/U_{0}$ values, the AFM dimers are oriented parallel to each other, similar to $S_{3}$, but the spins of zigzag stripe and dimer stripe sitting at a distance of $a_{m}$(next-nearest neighbout) are FM aligned, unlike in the $S_{3}$ phase.

More exotic phases tend to appear at smaller values of $U_0/t$ closer 
to the metal-insulator transition boundaries. We observed a novel 10-sublattice non-coplanar phase named 10-vertices polyhedron ($P_{10v}$), displayed in Fig.~\ref{fig2}(g), where 10 distinct non-coplanar spins are placed on a double-striped Wigner crystal. The orientation of the spins is shown on the left side of Fig.~\ref{fig2}(g). Note that the spin sets $\{1,2,3,4,5,6\}$ and $\{1,9,10,4,7,8\}$ lie perfectly in the green and red planes, respectively, each set making a little contorted hexagon (i.e. the angle between spins ``5 and 6", ``5 and 4", etc. is nearly $60^{\circ}$; the precise angle between the red and green plane depends on the parameter values). Interestingly, the stripe of the spins ``4 and 1'' (oriented opposite to each other) is an AFM collinear stripe and the other stripe is non-collinear with either spins ``7,8,9, and 10" or ``2,3,5, and 6" still having nearest-neighbour AFM dimers similar to the $S_{(1,2,3)}$ phases.

It is interesting to note that for the larger screening length, namely with robust long-range interactions, the non-coplanar phases are stabilized in a larger region of the phase diagram for both $n=1/2$ and $2/3$ fillings. For $n=2/3$, we also observed, near the $U_{0}/t=10$ region and at large $U_{1}/U_{0}$, a state with AFM hexagonal stripes made up of nearest-neighbour sites which are separated by the nearest-neighbour AFM dimers, named Hexagonal Stripes (HS). The HS state respects the rotational symmetry of the honeycomb lattice unlike the double-striped states. We believe the double-striped Wigner crystal states will be most relevant for experiments at filling $n=2/3$ for honeycomb moir\'e superlattices. Its existence can be indirectly confirmed by optical anisotropy measurements~\cite{Jin01}.   

Similarly to the $n=1/2$ case, we found a fully polarised FM metallic phase for $U_{1}/U_{0}=0$, but now a larger $U_{0}/t$ is required. This $n=2/3$ fully polarised FM phase is rapidly suppressed when including long-range interactions, contrary to the $n=1/2$ case. A large portion of the small $U_{1}/U_{0}$ phase diagram (for $U_{0}/t \ge 10 $) is covered by the non-collinear Vortices phase, details are shown in the supplementary~\cite{supple}.

We also estimated the metal-insulator transition line by calculating the single particle density of states (shown in the supplementary~\cite{supple}). We found that the low $U_{1}/U_{0}$ phases, namely the FM and Vortices phases are mainly metallic. On increasing the $U_{1}/U_{0}$, the FM and Vortices phases gradually develops a weak charge density wave (CDW) and a small gap at the chemical potential and eventually system transits into the insulating Stripe phases with much stronger double-striped Wigner crystallization. The thick-dashed black line in the phase diagrams (in low $U_{1}/U_{0}$ region) depicts the metal-insulator transition.

\subsection{\it{(c) n=1/3 results}} 

Figures~\ref{fig3}(a,b) show the $U_{1}/U_{0}$ vs $U_{0}/t$ phase diagrams for $n=1/3$, again for both $d=2a_{m}$ and $d=10a_{m}$. As for other fillings, we will discuss the dominant phases in the $U_{0}/t\ge 10$ region. 
In the large $U_1/U_0$ we found a Zigzag Wigner crystal state with collinear AFM ordering $(\textrm{AFM+WC})$ present in a robust portion of the phase diagram, see Fig.~\ref{fig3}(c). The representative state for the $\textrm{AFM+WC}$ phase, see Fig~\ref{fig3}(b), has Bravais lattice vectors $(3\hat{x},0)$ and $(0,\sqrt{3}\hat{y})$. Similar zigzag pattern states for filling $n=1/3$ in a honeycomb lattice, forming a charge density, were also found by minimizing the energy of only the interaction part of the Hamiltonian~\cite{YZhang01}. Intuitively, the robustness of the $\textrm{AFM+WC}$ phase suggests it may survive adding quantum fluctuations because it dominates a large portion of the phase diagram, and hence it may be present for real honeycomb moir\'e materials at filling $n=1/3$. In the opposite limit of $U_{1}=0$, we again notice a fully polarised metallic FM phase at large $U_{0}/t$, as at $n=1/2$ and $2/3$ (near $U_{0}/t=10$ we also found a spiral phase in a small region). Turning on the long-range interactions, the FM phase now manifests itself via various CDW arrangements, in order of increasing $U_{1}/U_{0}$, for details see ~\cite{supple}.

For intermediate $U_{1}/U_{0}$, we found exotic magnetic patterns. For example, the coplanar state with, again, strong zigzag charge density wave stripes but with spins pointing at $120^\circ$ with respect to each other inside the zigzag stripe. However, the nearby twin $120^\circ$ stripe is oriented along some arbitrary angle, see Fig.~\ref{fig3}(d), hence we name this state zigzag $120^\circ$-twin stripes ($120^\circ$-TS). Moreover, we also found a collinear state with long zigzag spin stripes aligned opposite to each other, see Fig.~\ref{fig3}(e). 
We believe that these complex states emerge in the large $U_{0}/t$ and intermediate $U_{1}/U_{0}$ region 
as a compromise between strong tendencies towards an AFM Wigner crystal at larger $U_{1}/U_{0}$ and 
ferromagnetism at smaller $U_{1}/U_{0}$. It is interesting to note that for $n=1/3$ we did not find any non-coplanar states, in contrast to the $n=1/2$ and $n=2/3$ phase diagrams.  This might be due to smaller higher order exchanges between localized spins (like the ring-exchange discussed for the $n=1/2$ case)  because of relatively larger distances between the half-filled sites of Zigzag Wigner crystal states.

\begin{figure*}[!t]
\hspace*{-0.5cm}
\vspace*{0cm}
\begin{overpic}[width=2.0\columnwidth]{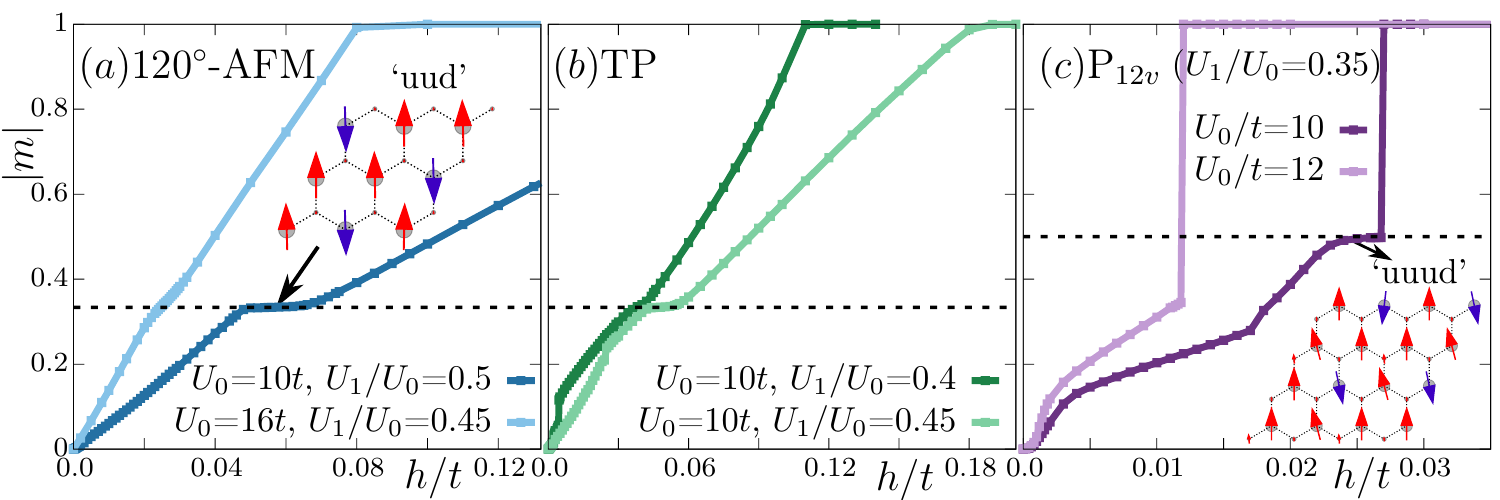}
\end{overpic}
\caption{Evolution of the net magnetization $|m|$ with an external magnetic field $h$ (along $\hat{z}$), at filling $n=1/2$ and $d=10a_{m}$. We selected representative parameter points for the respective phases using the phase diagram Fig.~\ref{fig1}(b). Panel (a) contains results for the points $(U_{0}=10t, U_{1}/U_{0}=0.5)$ and $(U_{0}=16t, U_{1}/U_{0}=0.45)$, choosen from the $120^{\circ}$-$\textrm{AFM}$ phase region. Panel (b) displays results for the Trigonal Prism (TP) state at $(U_{0}=10t, U_{1}/U_{0}=0.4)$ and $(U_{0}=10t, U_{1}/U_{0}=0.45)$. Results for the 12-vertices polyhedron state ($P_{12v}$) are shown in panel (c) at $(U_{0}=10t, U_{1}/U_{0}=0.35)$ and $(U_{0}=12t, U_{1}/U_{0}=0.35)$. The insets of panels (a) and (c) show the plateau states $uud$ ($|m|=1/3$) and $uuud$ ($|m|=1/2$), respectively.}
\label{fig4}
\end{figure*}

Our discovery of a variety of exotic spin arrangements close to the dominant AFM and FM states, including the $n=1/2$ and $n=2/3$ exotic phases, is common
occurrence in other families of materials where there is strong phase competition,
such as in manganites~\cite{manga1,manga2,manga3,manga4,manga5}, ruthenates~\cite{ruthe1},
and ladder iron superconductors~\cite{ladder1,ladder2}. These unusual states all arise because near
a possible first-order AFM-FM transition, there are spin arrangements mixing both characteristics of
the dominant states that can reduce even further the energy.

\section{Magnetization evolution as signature of exotic states}
Not all tools available to observe the magnetic structure of bulk
materials are available for two-dimensional moir\'e materials.  In particular neutron diffraction methods 
cannot be applied because of small samples sizes and magnetizations that are very small when measured per atom.
Fortunately, it turns out that the total spin magnetization of moir\'e materials can be measured optically \cite{Tang01}, at least in the $K$-valley case. With the primary purpose of assessing how optical studies might see the 
magnetization of moir\'e materials, in this section we will discuss the evolution of the net magnetization $|m|=2/N_{e}|\sum_{i}\langle S_{i}\rangle|$ with external magnetic field $h$ along the $\hat{z}$ direction,
which adds a Zeeman splitting term $-1/2\sum_{i}(n_{i\uparrow} -n_{i\downarrow})h$ to the Hamiltonian.

Firstly, we discuss the $n=1/2$ case (for $d=10a_{m}$), with main focus on the phases at large $U_{1}/U_{0}$, namely the $120$-$\textrm{AFM}$, Trigonal Prism (TP), and 12-vertices polyhedron ($P_{12v}$) phases. We use two representative parameter points [$(U_{0}=10t, U_{1}/U_{0}=0.5)$ and $(U_{0}=16t, U_{1}/U_{0}=0.45)$] for the $\textrm{120}$-$\textrm{AFM}$ state [see phase diagram in \ref{fig1}(b)]. The magnetization evolution for the $120$-$\textrm{AFM}$ state is shown in Fig.~\ref{fig4}(a). As $h/t\rightarrow 0$, $|m|$ increases linearly, $|m|\approx \chi_{s}h/t$.  We find that the spin susceptibility defined as $\chi_{s}=\lim_{h\to0} d|m|/dh$ is 
close to 8 and 16 for $(U_{0}=10t, U_{1}/U_{0}=0.5)$ and $(U_{0}=16t, U_{1}/U_{0}=0.45)$, respectively. 

Remarkably, at $(U_{0}=10t,~ U_{1}/U_{0}=0.5)$ a very clear $|m|=1/3$ plateau was unveiled, while at
$(U_{0}=16t, U_{1}/U_{0}=0.45)$ this was replaced by a small, but still visible, kink.
The plateau in magnetization signals the stability of a three-sublattice  
{\it uud} state, where $u(d)$ denotes up(down)-spin. This {\it uud} state is shown in the inset of 
Fig.~\ref{fig4}(a). The presence of a $|m|=1/3$ plateau in $|m|$ vs $h$ curves was reported before 
for a quantum spin $S=1/2$ nearest-neighbour antiferromagnetic Heisenberg model on a triangular lattice, 
employing exact calculations at zero temperature~\cite{Chu02};
the classical spin Heisenberg model does not show this plateau~\cite{Starykh01} at least for the nearest-neighbour interactions. 
Thus, it is fascinating that our HF calculations can capture this 
$|m|=1/3$ plateau in the emergent triangular lattice Wigner crystal, especially
at $(U_{0}=10t, U_{1}/U_{0}=0.5)$ which is close to the region of competition with non-coplanar phases. 
As already discussed, 
ring-exchange is expected to play an important role in non-coplanar phases. 
Thus, it is reasonable to invoke such terms in the effective spin model to explain 
the plateau's formation in a mean-field calculation. This conclusion is
in agreement with calculations for the {\it classical} spin models including ring-exchange which 
also shows plateau formation in $|m|$ vs $h$ curves~\cite{Momoi01,Momoi02}. 

We also investigated the evolution of $|m|$ with $h$ in non-coplanar phases [see Fig.~\ref{fig4}(b,c)]. We found that the Trigonal Prism phase shows a very similar $|m|$ vs $h$ curve as displayed before, i.e. with a linearly increasing $|m|$ for small magnetic field $h$ and the presence of a $|m|=1/3$ plateau. Interestingly, the 12-vertices polyhedron phase shows distinctively different features in the $|m|$ evolution [see Fig.~\ref{fig4}(c)], with  $\lim_{h\to0}|m|\approx \chi_{s}h + a h^2$ where $a/\chi_{s}>>1$. For example,  at $(U_{0}=10t, U_{1}/U_{0}=0.35)$ we found $\chi\approx0.82$ and $a\approx 5\times 10^4$. We believe that the small  susceptibility and robust second-order term arises from the presence of ferromagnetic exchange in the $P_{12v}$ phase: as discussed in ~\cite{Kubo01} the $P_{12v}$ phase can be obtained from the ferromagnetic classical spin model with ring-exchange on a triangular lattice. Moreover, we found the presence of a $|m|\approx 1/2$ plateau  with a slightly canted {\it uuud} state (see inset of Fig.~\ref{fig4}(c)) at location $(U_{0}=12t, U_{1}/U_{0}=0.35)$. We noticed that having a robust $|m|=1/2$ plateau depends on the parameter values we select, and was only present at large $U_{0}/t$ values. The magnetization evolution for the $P_{12v}$ phase always shows a first-order transition to the saturated ferromagnetic state, with a robust jump of $|\Delta m|\ge 1/2$ unlike the case of the $120$-$\textrm{AFM}$ and Trigonal Prism phases, where $|m|$ smoothly increases up to its saturation value with a gradual spin canting in the states.

The saturation magnetic field values can be converted in Teslas using $H=h/2\mu_{B}$ where $\mu_{B}$ is the Bohr magneton. Note that the hopping parameter $t$ increases with the twist angle ($\theta$) between the layers and varies in the approximate range $\{0.16,3\}$~meV for $\theta \in \{1.5^\circ,2.5^\circ\}$ (for details, see~\cite{supple}). Using these hopping values, for a given $U_{0}/t$ we can crudely estimate the range of saturation magnetic fields. For example, we checked that in the $120^\circ$ phase at $(U_{1}/U_{0}=0.5,U_{0}=10t)$ (the point with the largest saturation field $h/t=0.24$ in the phase diagram of Fig.~\ref{fig1}(b)) the saturation is reached at 0.35~T for $t=1/6$~meV and at 6.2~T for $t=3$~meV. This analysis suggests that saturation magnetic fields can easily be reached experimentally. We also studied the magnetization evolution of Wigner crystal phases for the fillings $n=2/3$ and $1/3$ but did not find any fractional magnetization plateaus (see ~\cite{supple}).

\section{Discussion}

In this work, we comprehensively studied the $U_{1}/U_{0}$ vs $U_{0}/t$ phase diagrams of a moir\'e
honeycomb lattice model with long-range Coulomb interactions, for fillings $n=1/2, 1/3$, and $2/3$, employing the Hartree Fock approximation. We believe this study, specially the larger $U_{1}/U_{0}$ regions of the phase diagrams
with Wigner crystal ground states, is directly relevant for potential future experiments on $\Gamma$-valley homobilayer-TMD's, such as $\textrm{MoSe}_{2}, \textrm{MoS}_{2}$, and $\textrm{WS}_{2}$ \cite{Angeli01,Xian01,Vitale01}. Specifically, we have found emergent patterns involving triangular lattices, double-striped, and zig-zag Wigner crystals for the fillings $n=1/2, 2/3,$ and $1/3$, respectively, in the region of robust long-range correlation strengths. These Wigner crystals can be directly observed using recently developed high-resolution sensitive, but non-invasive, scanning tunneling microscopy which has already been employed to image Wigner crystals in fractionally filled triangular $\textrm{WSe}_{2}/\textrm{WS}_{2}$ moir\'e superlattices~\cite{HLi01}. Moreover, indications for stripe and zig-zag charge ordering in Wigner crystals at $n=2/3$ and $1/3$, respectively -- states that break rotational symmetry -- can be indirectly observed via anisotropies in optical conductivity measurements~\cite{Jin01}.
 
We extensively studied the magnetism of Wigner crystal states and found that the long-range interactions can lead to an unexpected abundance of novel magnetic phases. As intuitive guidance to the vast complexity observed, we noticed common features at the three densities $n=1/2,1/3$, and $2/3$. For instance, near the 
metal insulator transition, the system always transits from a FM-metal 
to an AFM Wigner crystal insulator (at robust long-range electronic repulsion), often via an intermediate FM charge density wave and other more complex states (with $|m|=0$). These exotic states emerge from the competition between the robust FM vs AFM Wigner system tendencies in the small and large $U_1/U_0$ extremes of the phase diagram. Interestingly, we noticed for all three fillings ($1/2,2/3,$ and $ 1/3$) the fully polarized FM ($|m|=1$) to AFM ($|m|=0$) transition is of first order. In other words the magnetization drops from 1 to 0 suddenly on increasing the strength of $U_{1}/U_{0}$, at fixed $U_{0}/t$. 
As recently shown, the magnetic field induced Zeeman splitting in moir\'e excitons can be used to measure the magnetic susceptibility and the related Weiss constant, directly indicating whether the system is in any of the above discussed FM to AFM regions~\cite{Tang01,CJin02}.
 
In real experiments, tuning the distance between the gate and the device ($D$)~\cite{Huang01} directly affects the screening length $d$ as $d=2D$, which primarily corresponds to varying  the $U_{1}/U_{0}$ strength  of the long-range interaction. Similarly, changing the dielectric environment to tune the dielectric constant $\epsilon$ corresponds to moving along the horizontal axis of our phase diagrams i.e. 
tuning $U_{0}/t$ while keeping constant $U_{1}/U_{0}$ because
 $U_{i} \propto \epsilon^{-1}$ for any $i\ge 0$. Thus,  we believe that changing the screening length 
and dieletric enviroment can help to reach the most stable exotic non-collinear and non-coplanar phases discussed in this work. 

We also discussed the magnetization evolution with and external magnetic field, for the various exotic states, which can be used experimentally as indirect evidence for the novel states found here. Our study shows that honeycomb moir\'e materials can harbor very rich spin physics in two dimensions, as much as in the extensively studied triangular moir\'e materials. Interestingly, recent high resolution angle-resolved photoemission spectroscopy and scanning tunneling microscopy experiments have shown the presence of $\Gamma$-valley moir\'e bands and its associated real-space honeycomb lattice, respectively, in the twisted $\textrm{WSe}_{2}$ homobilayer~\cite{Pie01}. Moreover, ab-initio calculations showed that the sufficient pressure on twisted $\textrm{WSe}_{2}$ homobilayer  can push $\Gamma$-valley moir\'e bands to higher energy, mainly above the $K$-valley bands, making them the valence bands. Above discovery broadens the horizon of our work as the results discussed in this paper can also be realised in the already synthesized twisted $\textrm{WSe}_{2}$ homobilayer materials. We hope our study will also encourage experimentalists to synthesize the $\Gamma$-valley homobilayer TMDs because exotic magnetic states and Wigner crystals can potentially be achieved in these materials, as our work indicates. 

{\fontsize{8pt}{12pt}\selectfont


\section{{\fontsize{8pt}{10pt}\selectfont Methods}}\label{Method}
We will now describe the mean field method used in this work. We can imagine the $L_{x}\times L_{y}$ lattice tessellated by an $n_{x} \times n_{y}$ number of smaller cells of $l_{x}\times l_{y}$ size, where $l_{x(y)}=L_{x(y)}/n_{x(y)}$. In the basis of the smaller cells, we can write the Hamiltonian interaction terms of Eq.\ref{H_RealSpace} in the following manner:
\begin{multline}
H_{int} = U_{0}\sum_{j,\alpha,a}n_{j \alpha a\uparrow}n_{j \alpha a\downarrow} + U_{1}\sum_{j\alpha}n_{j\alpha 0}n_{j \alpha 1}\\ + 1/2\sum_{j, \alpha\ne\alpha^{\prime},a,b}U_{ab}^{\alpha \alpha^{\prime}}n_{j\alpha a}n_{j \alpha^{\prime} b} \\+ 1/2\sum_{j\ne j^{'},\alpha,\alpha^{\prime}, a, b}U_{ab}^{jj^{\prime}\alpha\alpha^{\prime}}n_{i\alpha a}n_{j\alpha^{\prime}b}.
\end{multline}
Now the site index $i$ of Eq.\ref{H_RealSpace} is replaced by $\{j,\alpha \}$, where $j$ is the cell index and $\alpha$ is the site index with respect to the $j$th cell. All four fermionic terms in the above Hamiltonian are treated under the Hartree-Fock approximation. The (many) order parameters to be used to minimize the energy are the elements of the single particle density matrix, namely  $\langle c_{j \alpha a \sigma}^{\dagger}c_{j^{\prime} \alpha^{\prime} b \sigma^{\prime}} \rangle$. We assume the mean-field solution can have broken translational symmetry with a new emergent unit cell $l_{x}\times l_{y}$. Under the above assumption we can write that any observable satisfies $O(j,\alpha,j^{\prime},\alpha^{\prime})=O(\mathbf{r}_{j^{\prime}}-\mathbf{r}_{j},\alpha^{\prime},\alpha)$. After imposing the above assumption on the order parameters, using $c_{j\alpha a \sigma}^{\dagger} = \frac{1}{\sqrt{N_{cells}}}\sum_{\mathbf{k}}e^{\iota\mathbf{k}.\mathbf{r}_{j}}c_{\mathbf{k} \alpha a \sigma}^{\dagger}$ we can write the Hamiltonian in momentum ($\mathbf{k}$) space as follows,
\begin{multline}
H_{int}^{HF}(\mathbf{k})=U_{0}\sum_{\alpha,a,\sigma}\langle n_{\alpha a \bar{\sigma}} \rangle n_{\mathbf{k}\alpha a \sigma} - (\langle S_{\alpha a}^{+} \rangle S_{\mathbf{k}\alpha a}^{-} + h.c.) \\ + U_{1}\sum_{\alpha, \sigma, \sigma^{\prime} ,a}\langle n_{\alpha \bar{a}\sigma}\rangle n_{\mathbf{k}\alpha a \sigma^{\prime}} - \langle c_{\alpha \bar{a}\sigma}^{\dagger}c_{\alpha a \sigma^{\prime}}\rangle  c_{\mathbf{k} \alpha a \sigma^{\prime}}^{\dagger}c_{\mathbf{k} \alpha \bar{a} \sigma} \\ +
\sum_{\substack {\alpha\ne\alpha^{\prime},\\ a,b,\sigma,\sigma^{\prime}}}U_{ab}^{\alpha\alpha^{\prime}}(\langle n_{\alpha a \sigma} \rangle n_{\mathbf{k}\alpha^{\prime} b \sigma^{\prime}} - \langle c_{\alpha a \sigma}^{\dagger} c_{\alpha^{\prime} b \sigma^{\prime}} \rangle  c_{\mathbf{k}\alpha^{\prime} b \sigma^{\prime}}^{\dagger}c_{\mathbf{k}\alpha a \sigma} ) \\ + \sum_{\substack {\alpha, \alpha^{\prime}, \\ a,b,\sigma, \sigma^{\prime}}} \bar{U}_{ab}^{\alpha \alpha^{\prime}} \langle n_{\alpha^{\prime} b \sigma^{\prime}}\rangle n_{\mathbf{k} \alpha a \sigma} - V_{\alpha a \sigma}^{\alpha^{\prime}b\sigma^{\prime}}(\mathbf{k})c_{\mathbf{k}\alpha^{\prime}b\sigma^{\prime}}^{\dagger}c_{\mathbf{k}\alpha a\sigma}
\end{multline} 
where $\bar{U}_{ab}^{\alpha\alpha^{\prime}}=\sum_{j\ne0}U_{ab}^{0\alpha,j\alpha^{\prime}}$, and $V_{\alpha a \sigma}^{\alpha^{\prime}b\sigma^{\prime}}(\mathbf{k})=\sum_{j\ne 0}U_{ab}^{0\alpha,j\alpha^{\prime}}e^{\iota\mathbf{k}j}\langle c_{0\alpha a\sigma}^{\dagger}c_{j\alpha^{\prime}b\sigma^{\prime}}\rangle$.
The kinetic energy term in $\mathbf{k}$-space becomes
\begin{equation}
H_{KE}(\mathbf{k}) = \sum_{\alpha, \alpha^{\prime}, a,b,\sigma, \sigma^{\prime}}\epsilon_{\alpha^{\prime}b\sigma^{'}}^{\alpha a \sigma}(\mathbf{k})c_{\mathbf{k} \alpha a \sigma}^{\dagger}c_{k\alpha^{'}b\sigma^{'}}
\end{equation}
where $\epsilon_{\alpha^{\prime}b\sigma^{'}}^{\alpha a \sigma}(\mathbf{k})=\sum_{j}t_{0\alpha^{\prime}b\sigma^{\prime}}^{j\alpha a \sigma} e^{\iota \mathbf{k}\cdot \mathbf{r}_{j}}$. The total Hartree-Fock Hamiltonian is then written as 
\begin{equation}
H^{HF} = \sum_{\mathbf{k}} H_{KE}(\mathbf{k}) + H_{int}^{HF}(\mathbf{k}) - 1/2\langle H_{int}^{HF}(\mathbf{k})\rangle.
\end{equation}

The main advantages of performing Hartree-Fock in this manner, for a given $l_{x(y)}<L_{x(y)}$, is two folded: (i) as $H^{HF}$ is block-diagonal in $\mathbf{k}$-space only the diagonalizations of smaller block matrices corresponding to different $\mathbf{k}$'s is needed, rendering the calculation faster; (ii) unrestricted Hartree-Fock calculations starting from random values tend to converge to inhomogenous states with higher energy than the actual mean field ground states found by our methodology. Here, by performing calculations for various clusters $l_{x} \times l_{y}$ and comparing energies, increases our confidence of convergence towards the true Hartree-Fock ground state. Obviously for $l_{x(y)}=L_{x(y)}$ the above technique becomes the canonical real-space unrestricted Hartree-Fock, which is also performed in the present work.

The Hartree-Fock ground state calculations were performed for system sizes up to $12 \times 12$. Self-consistent solutions with a maximum error of $10^{-5}$ were achieved starting from 10 different random initial configurations for the order parameters (while assuming the solution fits in $4 \times 4$, $6 \times 6$, and $12 \times 12$ emergent magnetic unit cells), which makes for a total of 30 runs for each parameter point. The converged solution with the lowest energy was used as the ground state. The number of different order parameters for the $l_{x}=l_{y}=12$ case are nearly $1.6 \times 10^{5}$, giving an idea of how challenging this effort becomes when searching for unbiased results within real-space Hartree Fock. To accelerate the convergence we used the Anderson mixing approach~\cite{Johnson01}.

}
\FloatBarrier


{\fontsize{8pt}{12pt}\selectfont


\section{{\fontsize{8pt}{10pt}\selectfont Acknowledgements}}\label{Acknow}
N.K. and E.D. were supported by the US Department
of Energy (DOE), Office of Science, Basic Energy Sciences (BES), Materials Science and Engineering Division.
N.M.D and A.H.M. were supported the U.S. Department of Energy, Office of Science, Basic Energy Sciences, under Award No. DE-SC0022106.}

\end{document}